\documentclass[prb,showpacs,amsmath,floatfix,twocolumn]{revtex4-1}

\usepackage{subfigure}
\usepackage{graphicx}
\usepackage{amssymb}

\begin{document}

\title{Size of stripe domains in a superconducting ferromagnet}
\author{Vu Hung Dao, S\'ebastien Burdin, and Alexandre Buzdin}
\affiliation{
Laboratoire Onde et Mati\`ere d'Aquitaine, UMR 5798 du CNRS,\\ 
Universit\'e Bordeaux 1, 351 cours de la Lib\'eration, 33405 Talence, France}
\date{March 10, 2011}

\begin{abstract}
In a superconducting ferromagnet, the superconducting state appears in the ferromagnetic phase where usually a domain structure  has already developed. We study the influence of the superconducting screening currents on a stripe structure with out-of-plane magnetization, in a film of arbitrary thickness. We find that superconductivity always induces a shrinkage of the domains, and there is a critical value of penetration depth below which a mono-domain structure is more stable than the periodic one. Furthermore we investigate the possible different effects of singlet and triplet superconductivity on the domain width, as well as the conditions for the existence of vortices in the domains. The obtained results are then discussed in light of the experimental data of superconducting ferromagnets URhGe, UGe2, and UCoGe.
\end{abstract}


\maketitle

\section{Introduction and model}

\begin{figure}
	\centering
	\scalebox{0.9}{
		\includegraphics{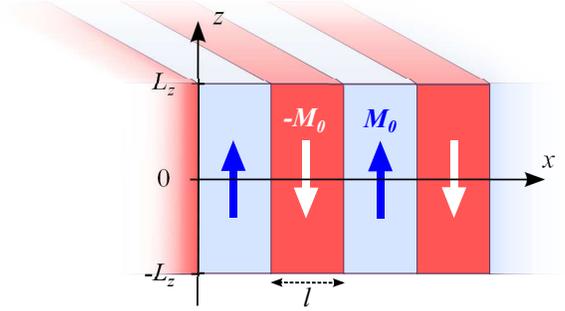}}
	\caption{(color online). Geometry of the considered stripe domain structure: the domain width is $l$ and the film thickness is $2 L_z$. The alternating magnetizations ${\bf M}$ of the domains are perpendicular to the film.}
	\label{fig:ferroDS}
\end{figure}

In the discovered Uranium-based superconducting ferromagnets (SFM) UGe$_2$~\cite{saxena}, URhGe~\cite{aoki2001} and UCoGe~\cite{huy} the Curie temperature $\Theta$ is much higher than the superconducting critical temperature $T_c$ which means that superconductivity appears in the ferromagnetic state where it is usually a domain structure (DS) which develops. In previous works~\cite{krey,bulaevskii,sonin1,faure,sonin2} the influence of the superconducting screening currents on the DS has been studied in the case of thick or bulk systems, when the thickness $2 L_z$ along the easy magnetization direction is much larger than both the transverse domain width $l$ (see Fig.~\ref{fig:ferroDS}) and the London penetration depth $\lambda$. The  domain period $2 l$ at equilibrium results from the balance between a positive contribution to the energy density due to domain walls and a negative contribution from the magnetic induction, as the magnitude of both increases when $l$ is reduced. An exact energy minimization~\cite{faure} shows that for the condition $\lambda > \tilde{w}/( 8 \pi)$ (where $\tilde{w}$ is an effective domain wall width that parametrizes the wall energy) superconductivity decreases the domain size due to partial penetration of the magnetic field near the domain wall. This energy decrease is proportional to $\lambda$ and the formation of the domain wall is favorable when this contribution counterbalances the energy of the domain wall itself, which is proportional to $\tilde{w}$. For $\lambda < \tilde{w}/( 8 \pi)$ the system is in a mono-domain state without any domain wall~\cite{faure,sonin1,sonin2}. In the present article we extend Faur\'e and Buzdin's work \cite{faure} by deriving the expression of the energy valid for any thickness $2 L_z$ and we discuss the DS in all limits, in particular when $\lambda, l \gg L_z$. Furthermore we investigate the possible  effects of singlet and triplet superconductivity on the DS, as well as the conditions for the existence of vortices in the domains. 

\emph{Model.\textemdash} We consider a ferromagnetic film of thickness $2 L_z$ that can also become superconducting (see Fig.~\ref{fig:ferroDS}). The $z$-axis is chosen perpendicular to the film with the surface edges at $z=\pm L_z$. Domain walls parallel to the $yz$-plane separate the periodic ferromagnetic structure into domains of equal width $l$ and with magnetization ${\mathbf M}= M(x) {\mathbf e}_z$ alternating along the $x$-axis, i.e.  
$ M(x)=  \pm M_0 = (4 M_0/l) \sum^{\infty}_{k=0} \sin (q x)/q$
with  
\begin{equation}
q \equiv \frac{(2k+1)\pi}{l}.
\end{equation}
This means that we consider the domain wall thickness very small compared to $l$ and $\lambda$. The energy per surface unit is $\mathcal{F}(\mathbf{B}, l)=\mathcal{F}_M(\mathbf{B}, l)+\mathcal{F}_{\textrm{SC}}(\mathbf{B}, l)+\mathcal{F}_{\textrm{DW}}(l)$ where the magnetic field energy is given by the relation 
\begin{equation}
\mathcal{F}_M (\mathbf{B}, l) =  \frac{1}{8\pi d_x d_y}  \left( \int_{\! |z|\! \leq \! L_z} \!\!\!\!\!\! |\mathbf{B} - 4\pi \mathbf{M}|^2  dV +  \int_{\! |z|\! >\! L_z} \!\!\!\!\!\!\!\!  |\mathbf{B}|^2  dV \right)~, 
\end{equation}
and the superconducting current energy is expressed in the London limit~\cite{degennes} (i.e. the superconducting coherence length $\xi \ll \lambda, l$)
\begin{equation}
\mathcal{F}_{\textrm{SC}} (\mathbf{B}, l) =  \frac{1}{8\pi d_x d_y}    \int_{\! |z|\! \leq \! L_z} \!\!\!\!\!\! \lambda^2 |\nabla \times(\mathbf{B} - 4\pi \mathbf{M})|^2  dV~.  
\end{equation}
Here, $d_x d_y$ is the total area the film surface. The explicit dependences with respect to the domain width result from the Ansatz we made for the magnetization $\mathbf{M}(l)$. 
The contribution from the domain walls may be written as~\cite{faure}
\begin{equation}
\mathcal{F}_{\textrm{DW}} (l) = \frac{M_0^2 \tilde{w} L_z}{l}~, 
\end{equation}
where $\tilde{w}$ is a domain wall characteristical length scale. Usually $\tilde{w}$ provides an upper limit of the real domain wall width~\cite{faure}. Hereafter, the temperature $T$ will be considered phenomenologically within an explicit dependence of the London penetration depth $\lambda(T)$. The latter is finite in the superconducting low$-T$ phase, and diverges at the critical temperature $T_c$. 

\emph{Outline of the article.\textemdash} In the second section  we analytically calculate the field distribution and the energy of a periodic stripe DS in a SFM of arbitrary thickness. An exact expression of the energy is obtained as an infinite sum. In section III the equilibrium size of the domains is determined in the superconducting state as a function of the penetration depth. Analytical expressions of the domain width are obtained in limit cases. The stability of the domain structure is also discussed. We then investigate the difference between singlet and triplet superconducting states. Specifically, the triplet pairing is described by a domain wall energy which is temperature dependent, in contrast with the singlet pairing. In section IV we draw the condition for vortices appearance at the center of a domain. Finally in section V we apply our results to the DS in Uranium-based superconducting ferromagnets before concluding. 

\section{Method of solution}
First, we minimize the energy $\mathcal{F}(\mathbf{B}, l)$ with respect to the magnetic induction $\mathbf{B}$. This yields the 
London equation $\Delta ({\mathbf B} - 4\pi{\mathbf M})= \lambda^{-2} {\mathbf B}$ in the film and $\Delta {\mathbf B}=0$ outside. Once $\mathbf{B}(l)$ is found by solving the London equation, the resulting total energy, $\mathcal{F}(l)=\mathcal{F}(\mathbf{B}(l), l)$, will be minimized with respect to $l$ to determine the domain width at equilibrium.

\subsection{Magnetic field $\mathbf{B}(l)$}

\begin{figure}
\begin{center}
\scalebox{0.72}{
 \includegraphics*{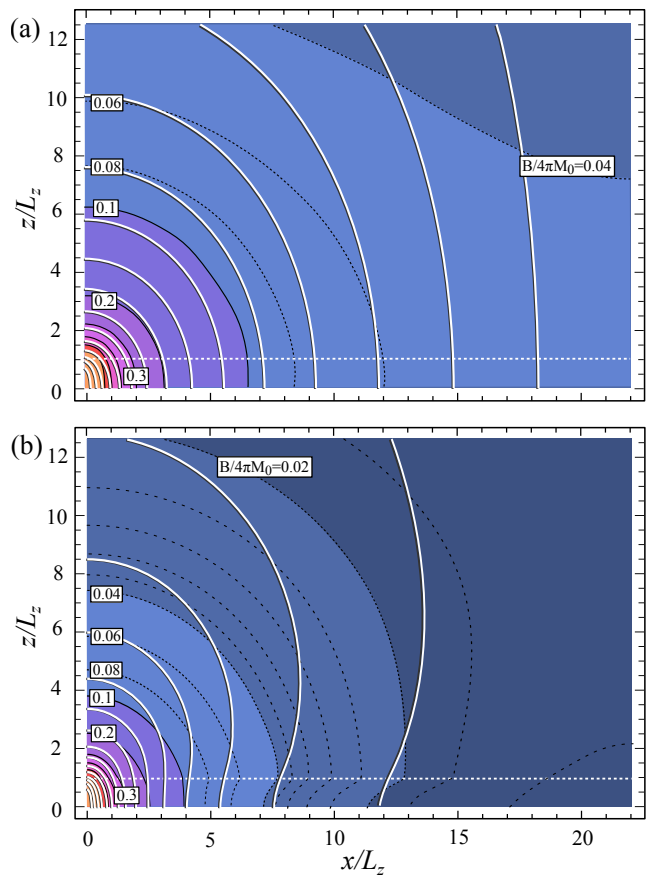}}
\end{center}
\caption{(color online). Magnetic field distribution of a wide domain ($l \gg L_z$) (a) in the normal state and (b) in the superconducting state ($\lambda=3L_z$). Flux lines (gray-and-white solid lines) and the contour plot of the field magnitude $|{\bf B}|/4\pi M_0$ (shaded background) are shown between $x=0$ and $x=l/2$. The horizontal dotted line is the upper film surface.}
\label{fig:mfield-a}
\end{figure}  

\begin{figure}
\begin{center}
\scalebox{0.72}{
 \includegraphics*{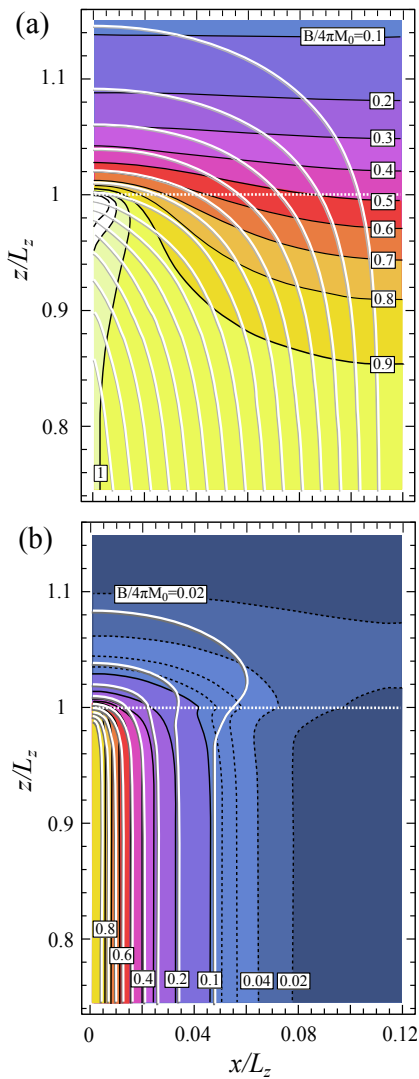}}
\end{center}
\caption{(color online). Magnetic field distribution of a narrow domain ($l \ll L_z$) (a) in the normal state and (b) in the superconducting state ($\lambda=0.02L_z$). Same conventions as in Fig.~\ref{fig:mfield-a}.}
\label{fig:mfield-b}
\end{figure}

We find the magnetic induction ${\mathbf B}$ by Fourier expansion. Using Maxwell-Thomson equation $\nabla \cdot {\mathbf B}=0$ and symmetry relations $B_z(-x)=-B_z(x)$ and $B_x(-x)=B_x(x)$, the field can be written as
\begin{equation}
{\mathbf B}(x,z)= \frac{16 \pi M_0}{l}
\sum_{k=0}^{\infty} \frac{\partial_z b_q(z)}{q} \cos (q x) {\mathbf e}_x +  b_q(z) \sin (q x) {\mathbf e}_z,
\end{equation}
where we remind that $q\equiv (2k+1)\pi/l$. When solving the London equation in the film and Maxwell's equation outside, we
use the symmetry condition $B_z(-z)=B_z(z)$ and the continuity condition at the surfaces. This yields, for $|z| \leq L_z$,
\begin{equation}
b_q(z)= \frac{q}{q_z^2}  \left( 1 - \frac{q  \cosh(q_z z)}{q_z \sinh (q_z L_z) + q \cosh(q_z L_z)} \right)~, 
\end{equation}
and for $|z| \geq L_z$, 
\begin{equation}
b_q(z)=\frac{q  \exp({-q(|z|-L_z)})}{q_z(q_z  + q \coth(q_z L_z))}~, 
\end{equation}
with 
\begin{equation}
q_z \equiv \sqrt{q^2 + \lambda^{-2}}.
\end{equation}
Figures~\ref{fig:mfield-a} and \ref{fig:mfield-b} show the distribution of the magnetic field in the normal state and the superconducting state. It is plotted between the domain wall and the domain center. In the normal state, the magnetic field distribution in a wide domain ($l \gg L_z$) is concentrated around the domain wall (see Fig.~\ref{fig:mfield-a}(a)) while in a narrow domain ($l \ll L_z$) it is nearly uniform and equal to $\pm 4\pi M_0 {\bf e}_z$ (see Fig.~\ref{fig:mfield-b}(a)). The supercurrent screens the field around the domain wall on a length scale $\lambda$ (compare for example the contour $|{\bf B}|/4\pi M_0=0.4$ in (a) and (b) of Fig.~\ref{fig:mfield-b}). Note that the supercurrent is responsible for a kink that the lines of constant $|{\bf B}|$ show at the film surface (see Fig.~\ref{fig:mfield-a}(b) and Fig.~\ref{fig:mfield-b}(b)). This is because $\nabla \times {\bf B}$ is discontinuous at the interface when $\lambda^{-2} \neq 0$. Flux lines also have a kink at the domain wall since the magnetization ${\bf M}$ is discontinuous there.

\subsection{Energy $\mathcal{F}(l)$ of the domain structure}
To simplify the expressions of the energy it is convenient to introduce the normalized lengths 
\begin{equation}
L\equiv\frac{l}{L_z},\;\; \Lambda\equiv\frac{\lambda}{L_z},\;\; \Lambda_{\rm eff}\equiv\frac{\lambda_{\rm eff}}{L_z} \;\;\textrm{and}\;\; W\equiv\frac{\tilde{w}}{L_z}~,
\end{equation}
where $\lambda_{\textrm{eff}}\equiv\lambda^2/L_z$ is Pearl's penetration depth which, in the limit $L_z \ll \lambda$, takes the place of $\lambda$ as the effective magnetic length scale\cite{pearl}. Note that in these notations $\Lambda_{\rm eff}= \Lambda^2$.  
Furthermore we will consider the normalized energy $\bar{\mathcal{F}}= \mathcal{F}/(32 \pi M_0^2 L_z)$ and omit the bar henceforth. So the domain wall contribution is $\mathcal{F}_{\textrm{DW}}(L) = W/(32 \pi L)$ and
the volume contribution, $\mathcal{F}_{\text{vol}}\equiv \mathcal{F}_M + \mathcal{F}_{\textrm{SC}}$ is
\begin{equation}
\mathcal{F}_{\text{vol}} (L) =  \frac{1}{8} - \frac{\Lambda}{4L} \tanh{\frac{L}{2\Lambda}} +  \frac{1}{L^2} \sum_{k=0}^{\infty} \frac{Q}{Q_z^3 (Q_z + Q \coth Q_z)}~, 
\label{eq:Fvol}
\end{equation}
where $Q \equiv (2k+1)\pi/L$ and $Q_z \equiv \sqrt{\Lambda^{-2} + Q^2}$. For a thick film, i.e. $Q_z \equiv q_z L_z \gg 1$ so that $\coth (Q_z)=1$, Faur\'e and Buzdin's result~\cite{faure} is recovered as expected. Since the limit $l \ll L_z$ has been previously investigated~\cite{faure} we discuss below the limit $l \gg L_z$, that is $L \gg 1$, with more details.

\subsubsection{Regime $\lambda \gg l$}
This parameter regime is realized in the normal phase where $\lambda=\infty$, as well as inside the superconducting phase, in the vicinity of the critical temperature $T_c$. 
When $\lambda\gg l$ (i.e., when $\Lambda\gg L$), the sum in (\ref{eq:Fvol}) may be expanded in powers of $L/\Lambda$ before summing. Hence for $\lambda \gg l \gg L_z$, 
\begin{equation}
\mathcal{F}_{\textrm{vol}} (L)\approx  \frac{1}{8} + \frac{1}{8 \Lambda_{\rm eff}}  - \left( \frac{1}{2\pi} + \frac{2}{3\pi \Lambda_{\rm eff}} \right) \frac{\ln L}{L} + \frac{2 \ln \pi - 3}{4\pi L},
\label{eq:Fvol3}
\end{equation}
while for $l\ll L_z$,
\begin{equation}
\mathcal{F}_{\textrm{vol}} (L)\approx \frac{7 \zeta (3) L}{16 \pi^3} + \frac{L^2}{96 \Lambda^2}.
\label{eq:Fvol3b}
\end{equation}

\subsubsection{Regime $l \gg L_z$ and $l \gg \lambda_{\rm eff}$}
This parameter regime may be realized in the superconducting phase, for either thin films or large domains. 
When $L \gg 1$ the sum in (\ref{eq:Fvol}) can be approximated by an integral. Euler-MacLaurin's approximation yields
\begin{equation}
\mathcal{F}_{\textrm{vol}} (L)\approx \frac{1}{8} + \left( \frac{I(\Lambda)}{2\pi} - \frac{\Lambda}{4} \right) L^{-1} + \frac{\pi \Lambda^4}{12} L^{-3}
~, 
\end{equation}
where the $I(\Lambda)$ is defined by Eq.~(\ref{eq:I}) in the Appendix. 
Then using asymptotic expression (\ref{eq:F201}) of $I(\Lambda)$ in the limit $\Lambda \gg 1$, one finds 
\begin{equation}
\mathcal{F}_{\textrm{vol}} (L)\approx \frac{1}{8} + \frac{1}{\pi} \left( -\ln \Lambda + \frac{\ln 2}{2} - \frac{11}{24}\right) L^{-1} + \frac{\pi \Lambda^4}{12} L^{-3},
\end{equation} 
and for $\Lambda \ll 1$, asymptotic expression (\ref{eq:F202}) yields
\begin{equation}
\mathcal{F}_{\textrm{vol}} (L)\approx \frac{1}{8} +  \left( - \frac{\Lambda}{4} + \frac{(1-\ln 2)\Lambda^2}{2\pi}\right) L^{-1} + \frac{\pi \Lambda^4}{12} L^{-3}.
\end{equation}

\section{Domain width at equilibrium}
\label{sec:width}

\subsection{General results}

\begin{figure}
\begin{center}
\scalebox{1}{
 \includegraphics*{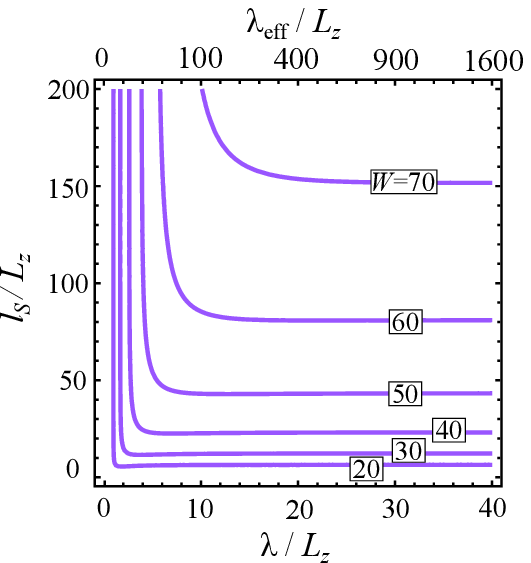}}
\end{center}
\caption{(color online). Normalized domain width $l_S/L_z$ as a function of the normalized penetration depth $\lambda/L_z$ for different values of the normalized effective domain wall thickness $W\equiv \tilde{w}/L_z$.}
\label{fig:ls01}
\end{figure}

\begin{figure}
\begin{center}
\scalebox{1}{
 \includegraphics*{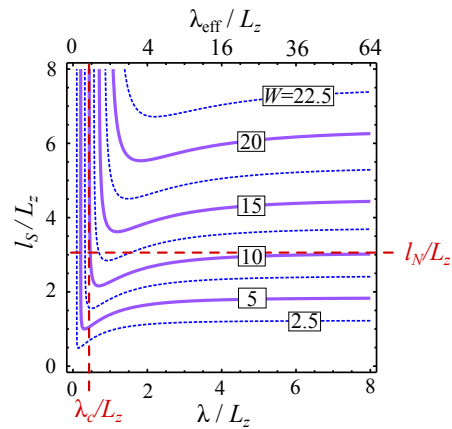}}
\end{center}
\caption{(color online). Normalized domain width $l_S/L_z$ as a function of the normalized penetration depth $\lambda/L_z$ for different values of the normalized effective domain wall thickness $W$. $l_S$ converges to the normal-state width $l_N$ when $\lambda$ tends to infinity, and it diverges at the critical value $\lambda_c$. $l_N$ and $\lambda_c$ are increasing functions of $W$.}
\label{fig:ls02}
\end{figure}

\begin{figure}
\begin{center}
\scalebox{1}{
 \includegraphics*{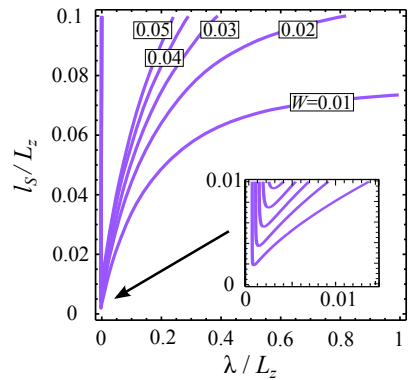}}
\end{center}
\caption{(color online). Normalized domain width $l_S/L_z$ as a function of the normalized penetration depth $\lambda/L_z$ for different values of the normalized effective domain wall thickness $W$.}
\label{fig:ls03}
\end{figure}

The equilibrium size $l_S$ is obtained by minimization of the total energy   $\mathcal{F}(l)=\mathcal{F}_{\text{DW}}(l)+\mathcal{F}_{\text{vol}}(l)$. Figures~\ref{fig:ls01}-\ref{fig:ls03} show $l_S$ as a function of the penetration depth $\lambda$ for different values of the normalized wall thickness $W$. In the superconducting state $\lambda$ decreases with decreasing temperature from infinity at $T=T_c$ to a finite value at $T=0$. As discussed below, just below $T_c$ the domain width always decreases from the normal state value $l_N$ realized for $\lambda=\infty$. One can distinguish two regimes. When the domains in the normal state are wide ($l_N \gg L_z$) the decrease is negligible, while when the domains are narrow ($l_N \ll L_z$) their width can drastically shrink down to $l_S \sim \lambda$. Then, in both regimes, $l_S$ reaches a minimum value before diverging at a critical lower bound $\lambda_c$. This limit corresponds to the situation where the ferromagnetic induction is completely screened by the supercurrent \cite{faure,sonin1,sonin2} and there is no stray field that need to be accommodated by a DS. 

The $\lambda$-dependence of the width is similar to the prediction for DS in superconductor-ferromagnet (S/F) bilayers \cite{bilayer,stankiewicz} except for two aspects. First the shrinkage factor in a S/F hybrid is limited and $l_S \geq \sqrt{2/3} l_N$ \cite{bilayer}. Experiments performed on heterostructures made of a ferromagnetic garnet layer combined with a superconducting layer of Pb~\cite{tamegai} or Nb~\cite{vlasko-vlasov} have observed the shrinkage of the domain size due to superconductivity. The smallest observed shrinkage factor $\sim 0.3$ was substantially smaller than the theoretical prediction $\sqrt{2/3}$. This circumstance is a non-equilibrium effect and is related to the special dynamics of the vortex structure created in the superconducting layer and coupled to ferromagnetic domains~\cite{vlasko-vlasov}. The second point is that although the DS can also be unstable in S/F  bilayers when $\lambda$ decreases, it exists a range of parameters (when the thickness of the F-layer is approximately larger than half the domain width in the normal state) in which $l_S$ remains finite even in the limit $\lambda=0$~\cite{stankiewicz}, in contrast with SFM where the DS is never stable for a vanishing $\lambda$.

Hereafter, complementary to the exact numerical solution, we present the analytical results obtained in various asymptotic regimes. 

\subsubsection{Normal state and vicinity of the transition}

\begin{figure}
\begin{center}
\scalebox{.82}{
 \includegraphics*{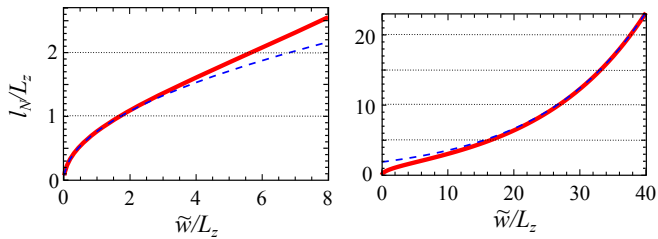}}
\end{center}
\caption{(color online). Normalized domain width $l_N/L_z$ in the normal state as a function of the normalized effective domain wall thickness $\tilde{w}/L_z$. Dashed lines are the plots of the asymptotic formulas.}
\label{fig:ln_w}
\end{figure}

In the limit $L \gg 1$, using expression (\ref{eq:Fvol3}) valid for the vicinity of the superconducting transition where $\Lambda^2 \gg L$, one finds that the minimum of the energy $\mathcal{F}(L)$ is realized at
\begin{equation}
L_S = \exp \left( 1 + \frac{W/16 - 3/2 + \ln \pi}{1 + 4/3 \Lambda^2} \right)~,
\end{equation}
which yields the domain width
\begin{equation}
L_N = \frac{\pi}{\sqrt{e}} \exp \left( W/16  \right)
\end{equation}
 in the normal state (see Fig.~\ref{fig:ln_w}). The domain width $L_S$ just below the superconducting transition is then related to the normal-state width $L_N$ by
\begin{equation}
L_S = L_N \left( 1- \frac{4 (\ln L_N - 1 )}{3 \Lambda^2} \right).
\end{equation}

In the other limit $L \ll 1$, the approximation (\ref{eq:Fvol3b}) yields the standard result~\cite{kittel,landau}
\begin{equation}
L_N= \sqrt{\frac{\pi^2}{14\zeta(3)} W},
\end{equation} 
(see Fig.~\ref{fig:ln_w}) and 
\begin{equation}
L_S = L_N \left( 1 - \frac{\pi^3}{42\zeta(3)} \frac{L_N}{\Lambda^2} \right)~,
\end{equation}
in accordance with Ref.~\cite{faure}. Hence the domain width always decreases when the system enters the superconducting phase.

\subsubsection{Stability of the periodic domain structure}

\begin{figure}
\begin{center}
\scalebox{.82}{
 \includegraphics*{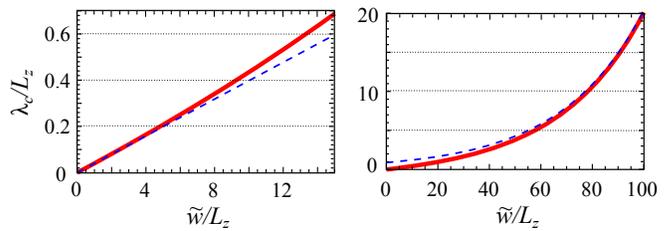}}
\end{center}
\caption{(color online). Normalized critical penetration depth $\lambda_c/L_z$ as a function of the normalized effective domain wall thickness $\tilde{w}/L_z$. Dashed lines are the plots of the asymptotic formulas.}
\label{fig:lambdac_w}
\end{figure}

\begin{figure}
\begin{center}
\scalebox{.79}{
 \includegraphics*{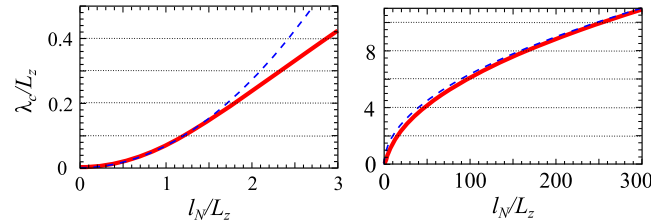}}
\end{center}
\caption{(color online). Normalized critical penetration depth $\lambda_c/L_z$ as a function of the normal domain width $l_N/L_z$. Dashed lines are the plots of the asymptotic formulas.}
\label{fig:lambdac_ln}
\end{figure}

The periodic domain structure is always stable in the normal state where the energy minimum is obtained for a finite width $L$. This is not the case in the superconducting state. In the limit $L \gg \Lambda^2$ and $L \gg 1$, the saddle-point equation for the energy is
\begin{equation}
L^2 =\frac{ \pi^2 \Lambda^4}{ \pi \Lambda - 2 I(\Lambda) - W/8}~,
\end{equation}
where the function $I(\Lambda)$ is defined by Eq.~(\ref{eq:I}) in Appendix. There is an energy minimum at a finite $L$ only for $\pi \Lambda - 2 I(\Lambda) - W/8 > 0$. Since $(\pi \Lambda - 2 I(\Lambda))$ increases with $\Lambda$, this condition defines a lower bound $\Lambda_c$ below which the periodic structure is unstable. In the thought experiment where $\Lambda$ decreases from infinity (normal state) to zero, the period $L_S$ first decreases from $L_N$ but then it increases back before diverging as $(\Lambda-\Lambda_c)^{-1/2}$ (see Fig.~\ref{fig:ls01}). Note that the limit of vanishingly small $\Lambda$ has been considered in Ref.~\cite{sonin1} and the conclusion of the absence of DS was made. When $L_N \gg 1$ in the normal state, the approximations of $I(\Lambda)$ yield 
\begin{equation}
\Lambda_c  \approx \sqrt{2} \exp{\left( \frac{W}{32} - \frac{11}{24} \right) } \approx \sqrt{\frac{2 L_N}{\pi e^{5/12}} }~, 
\end{equation}
while when the normal-state width $L_N \ll 1$, the periodic structure exists down to 
\begin{equation}
\Lambda_c \approx \frac{W}{8\pi} \approx \frac{7 \zeta(3)}{ 4 \pi^3} L_N^2.
\end{equation}
Figures~\ref{fig:lambdac_w} and \ref{fig:lambdac_ln} show that these analytical relations fit well the numerical results in the limit regimes.

One can now understand the difference of shrinkage factor between a narrow domain and a wide domain. For a narrow domain with $l$ and $\lambda \ll L_z$, the width in the superconducting state decreases with the penetration depth as $l_S \sim \lambda$. The shrinkage goes on until the DS is destabilized by an excessive superconducting screening. The minimum value that the width can have is then $l_S^{\rm min} \sim \lambda_c \sim l_N^2/L_z$, which is much smaller than $l_N$. In the opposite limit when $l$ and $\lambda \gg L_z$, the actual magnetic length scale is $\lambda_{\rm eff}$, not $\lambda$. And the domain width behaves as $l_S \sim \lambda_{\rm eff}$ when $\lambda_{\rm eff}$ is smaller than $l_N$. Thus the minimum $l_S^{\rm min} \sim \lambda_c^2/L_z \sim l_N $, so that a wide domain does not significantly shrink while the penetration depth decreases to the instability value $\lambda_c$. 

\subsection{The case of triplet superconductivity}

As we will see in a next section, the temperature dependence of the domain equilibrium size can provide a phenomenological way of distinguishing triplet from singlet pairing. Here, we describe the general formalism; numerical applications to some specific compounds will be shown later as examples, in the framework of Uranium-based superconductors. 

\subsubsection{Modification of the wall energy due to the depletion of condensation energy}
In the case of a triplet superconductor with Cooper pairs fully spin polarized by the intrinsic magnetization of the ferromagnetic domains, a depletion of the order parameter occurs on a width $\sim \xi$ at both sides of the interface separating two domains of opposite magnetizations. This inhomogeneity results in a loss of condensation energy $\sim 2 {\cal E}_{\rm{cond}} \xi d_y L_z$ at a domain wall.  At a first approximation the domain wall energy for $T \leq T_c$ is modified by replacing the effective domain wall width $\tilde{w}$ by the temperature dependent $\tilde{w}_t$ defined by

\begin{equation}
\tilde{w}_t \equiv \tilde{w} \left( 1 + \omega (1-t)^{3/2} \right),
\end{equation}
where $t\equiv T/T_c$, and
\begin{equation}
\omega \equiv \frac{2 {\cal E}_{\rm{cond}}(0) \xi(0)}{M_0^2 \tilde{w}}
\label{eq:omega}
\end{equation}
is proportional to the ratio of the superconducting energy at $T=0$ to the magnetization energy. Here, we have assumed $\xi(t)=\xi(0)/\sqrt{1-t}$, $\lambda(t)= \lambda(0)/\sqrt{1-t}$, and ${\cal E}_{\rm{cond}}(t) = {\cal E}_{\rm{cond}}(0) (1-t)^2$.
Using the asymptotic expressions relating $\tilde{w}$ to the normal state width $l_N$, one can estimate $\omega$ from experimental data in the limit $l_N \ll L_z$ with
\begin{equation}
\omega = \frac{ \pi^2 \xi(0) L_z {\cal E}_{\rm{cond}}(0) }{7 \zeta(3) l_N^2 M_0^2 },
 \end{equation}
and in the limit $l_N \gg L_z$ with
\begin{equation}
\omega = \frac{\xi(0) {\cal E}_{\rm{cond}}(0)}{ 8 \left( \ln  \frac{l_N}{\pi L_z} + \frac{1}{2} \right) L_z M_0^2}. 
\end{equation}

\subsubsection{Behavior of the domain width below the transition}

\begin{figure}
\begin{center}
\scalebox{.65}{
 \includegraphics*{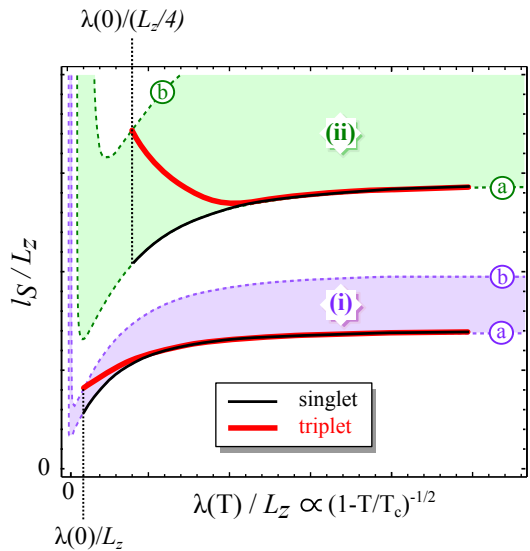}}
\end{center}
\caption{(color online). Schematic temperature dependence of the domain width $l_S$ for the singlet case (thin solid lines) and for the triplet case (thick solid lines) in (i) a film of thickness $L_z$ and (ii) in another one with a smaller thickness $L_z/4$. The dashed-line curves show the full $\lambda$-dependence of $l_S$ in the case of a constant effective wall thickness (a) $\tilde{w}=\tilde{w}_{t=1}$ and (b) $\tilde{w}=\tilde{w}_{t=0}$.}
\label{fig:ls_t}
\end{figure}

A triplet order parameter yields a vanishing correction to the domain wall energy at the transition. The temperature dependence of the domain size is in the limit $L_N \ll 1$,
\begin{equation}
L_S= L_N \left( 1 - \frac{\pi^3}{42 \zeta(3)} \frac{L_N}{\Lambda(0)^2} (1-t) + \frac{\omega}{2} (1-t)^{3/2} \right),
\end{equation}
and in the limit $L_N \gg 1$,
\begin{equation}
L_S= L_N \left( 1 - \frac{4 (\ln L_N -1)}{3 \Lambda(0)^2} (1-t) + \frac{\omega W}{16} (1-t)^{3/2} \right).
\label{eq:ls_triplet}
\end{equation}
So, as in the singlet case, $l_S$ always decreases with decreasing temperature in the vicinity of $T_c$. However at a lower temperature and for a large parameter $\omega$, the behavior for the triplet case can be significantly different from the singlet case. Since for a given $L_N$ there is an universal curve $L_S (\Lambda)$ for the singlet pairing, it is possible to detect an experimental signature of triplet superconductivity as a deviation from this curve. This is schematically illustrated in Fig.~\ref{fig:ls_t} which shows the temperature dependence of $L_S$ in the singlet and in the triplet case.  When temperature is reduced from $T_c$ (on the right side where $\lambda(T) \rightarrow \infty$) to $T=0$ (on the left side where $\lambda(T)=\lambda(0)$), the $L_S$-curve in the triplet case  interpolates between the curve (a) obtained with a constant $\tilde{w}=\tilde{w}_{t=1}$  and the curve (b) for $\tilde{w}=\tilde{w}_{t=0}$. The triplet-case curve follows the singlet-case curve with $\tilde{w}=\tilde{w}_{t=1}$ at high temperature but at a lower temperature a deviation appears which is maximum at $T=0$. 

In the expression~(\ref{eq:ls_triplet}) of $L_S$ for the triplet pairing, the first-order term is proportional to $L_z^2$ while the second-order term is proportional to $L_z^{-2}$ so one can expect that the singlet/triplet discrepancy can be amplified when reducing the thickness of the film. As shown in Fig.~\ref{fig:ls_t} this discrepancy is enhanced when the ratios $\lambda(0)/L_z$ and $\tilde{w}_t/L_z$ become larger. For small enough thickness the difference is not only quantitative but is also qualitative as the triplet-pairing curve can increase with decreasing temperature while the singlet-pairing curve decreases.

\section{Condition for vortex appearance}
\label{sec:vortex}
In this section we estimate the energy for the creation of a vortex-antivortex pair in the middle of adjacent domains. This additional energy is the sum of (i) the energy decrease when (anti)vortices are driven by screening currents away from domain boundaries to domain centers and (ii) the concomitant increase of the interaction energy between one vortex and its antivortex as they move apart from each other. For simplicity we neglect the vortex-core energy assuming the limit of large $\kappa=\lambda/\xi$.

\subsection{Energy decrease due to the screening current}
The force of the screening current acting on a vortex~\cite{degennes} is
$\frac{\Phi_0}{4\pi} \int_{-L_z}^{L_z} \left[ \nabla \times ({\bf b} - 4 \pi {\bf M}) \right] \times {\bf e}_z \; dz $, where the quantum of flux $\Phi_0= 2.07 \times 10^{-15}$ T.m$^2$. When the vortex moves from the domain wall to the domain center, the current then produces a work equal to
\begin{equation}
{\cal W}_{sc} = \frac{4 {\cal W}_{\infty}}{\Lambda^2 L} \sum_{k=0}^{\infty} \frac{(-1)^k}{Q_z^2} \left( \frac{1}{Q} - \frac{1}{Q_z(Q_z + Q \coth Q_z)} \right), 
\end{equation}
where ${\cal W}_{\infty} \equiv 2 L_z \Phi_0 M_0$ is the coupling energy of the flux carried by one vortex with the magnetization $M_0$ of a ferromagnetic domain. 

The work has simple analytical expressions in limit cases. In the regime $l \gg \lambda$,
\begin{equation}
{\cal W}_{sc} \approx {\cal W}_{\infty} \left( \! 1 - \frac{2}{\pi} \Phi_L \! \left( \! -1,1,\frac{1+L_z l/\pi \lambda^2}{2} \! \right) \! \right)~, 
\label{eq:WL5}
\end{equation}
where the Lerch transcendent $\Phi_L(z,s,a) \equiv \sum_{k=0}^{\infty} z^k/(a+k)^s$. If furthermore $l \gg  \lambda^2/L_z$, expression (\ref{eq:WL5}) can be simplified as
\begin{equation}
{\cal W}_{sc} \approx {\cal W}_{\infty} \left( 1 - \frac{2 \lambda^2}{L_z l} \right). 
\end{equation}
Thus ${\cal W}_{sc}$ increases with the width of the domain and, when the latter becomes large compared to the magnetic radius of a vortex, the work is simply equal to the coupling energy ${\cal W}_{\infty} \equiv 2 L_z \Phi_0 M_0$.

In the opposite regime $\lambda \gg l$, the screening current is vanishingly small so its work is only equal to a fraction of ${\cal W}_{\infty}$. For wide domains ($l \gg L_z$), the dependence of ${\cal W}_{sc}$ on the domain width is linear as 
\begin{equation}
{\cal W}_{sc}  \approx \frac{4 G {\cal W}_{\infty}}{\pi^2} \frac{l}{\lambda_{\rm eff}}, 
\end{equation}
where Catalan's constant $G \approx 0.916$, while for narrow domains ($l \ll L_z$), the dependence is quadratic as 
\begin{equation}
{\cal W}_{sc}  \approx \frac{{\cal W}_{\infty}}{8}  \frac{l^2}{\lambda^2} \label{eq:WL4}.  
\end{equation}

\subsection{Interaction energy of a vortex-antivortex chain}
The interaction energy of a chain composed by one vortex (antivortex) located at the center of every domain of positive (negative) magnetization is the magnetic energy required to separate by a distance $l$ vortices and their paired antivortices created at domain walls. In the case $\lambda \gg l$ or $\lambda_{\rm eff} \gg l$, the energy per one vortex~\cite{buzdin}
has the asymptotic expression
\begin{equation}
{\cal E}_{vv}= 2 L_z \left( \frac{\Phi_0}{4 \pi \lambda} \right)^2 \ln \frac{l}{\xi}.
\end{equation}
We have assumed $\xi \ll l$. In a large-$\kappa$ superconductor with $\xi(0) \ll l$, there is always a temperature regime close to $T_c$ in which $\lambda \gg l \gg \xi$. 

In the opposite limit, for $l \gg \lambda_{\rm eff} \gg L_z$ the interaction energy per vortex is
\begin{equation}
{\cal E}_{vv}= 2 L_z \left( \frac{\Phi_0}{4 \pi \lambda} \right)^2 \ln \frac{\lambda_{\rm eff}}{\xi},
\end{equation}
while for $L_z \gg l \gg \lambda$, it is
\begin{equation}
{\cal E}_{vv}= 2 L_z \left( \frac{\Phi_0}{4 \pi \lambda} \right)^2 \ln \frac{\lambda}{\xi}.
\end{equation}

\subsection{Condition for the appearance}
From the above results we can derive the condition of the vortex stability in a superconducting domain structure. The latter is assured when the total energy per vortex ${\cal E}_{\rm tot}= {\cal E}_{vv} - {\cal W}_{sc}$ is negative. This condition defines  a critical value of magnetization $M_v$ above which vortices appear. One can distinguish four regimes in which $M_v$ has a simple analytical expression. In a film with narrow domains (i.e. $l \ll L_z$), 
\begin{equation}
 4 \pi M_v = \frac{\Phi_0}{4\pi \lambda^2} \ln \!\! \left( \! \frac{\lambda}{\xi} \!\right) {\rm \; when \;} l \gg \lambda \label{eq:mc_1},
\end{equation} 
and
\begin{equation}
 4 \pi M_v =  \frac{2 \Phi_0 }{\pi l^2}  \ln \!\! \left(  \frac{l}{\xi} \right) {\rm \; when \;} l \ll \lambda \label{eq:mc_2}.
\end{equation}
Note that Eq.~(\ref{eq:mc_1}) recovers the expected result that vortices appear when the magnetization $4\pi M_0$ exceeds the lower critical field of a non-magnetic superconductor defined as $H_{c1}^*\equiv \frac{\Phi_0}{4\pi \lambda^2} \ln \!\! \left( \! \frac{\lambda}{\xi} \!\right)$. However, when the penetration depth is larger than the domain width, the coupling of the vortex flux with the DS magnetization is not energetically optimum so the critical magnetization (\ref{eq:mc_2}) is higher than the mono-domain value $H_{c1}^*$. 

In the opposite limit of wide domains (i.e. $l \gg L_z$), the critical magnetization is 
\begin{equation}
4 \pi M_v = \frac{ \Phi_0}{4 \pi \lambda^2} \ln \!\! \left( \! \frac{\lambda_{\rm eff}}{\xi} \!\right) {\rm \; for \;} l \gg \lambda_{\rm eff} \label{eq:mc_3},
\end{equation} 
while
\begin{equation}
4 \pi M_v = \frac{\pi \Phi_0}{16 G L_z l} \ln \!\! \left(  \frac{l}{\xi} \right)  {\rm \; for \;} l \ll \lambda_{\rm eff}  \label{eq:mc_4}.
\end{equation}
In this regime of parameters, $\ln{(  \lambda_{\rm eff}/\xi )} \gtrsim \ln{( \lambda/\xi )}$ so Eqs.~(\ref{eq:mc_3}) and (\ref{eq:mc_4}) can be interpreted as an extension of results Eqs.~(\ref{eq:mc_1}) and (\ref{eq:mc_2}) for the limit $l \gg L_z$ where $\lambda$ is replaced by $\lambda_{\rm eff}$.

\section{Application to Uranium-based superconducting ferromagnets}

\begin{widetext}

\begin{table}[ht]
 \centering
		\begin{tabular}{|c|| c|c|c|c|c||c|c|c|}
			\hline
			 & $2 \pi M_0^2$ (J/m$^3$) & $\xi(0)$ (nm) & $\lambda(0)$ ($\mu$m) & $L_z$ (mm) & $l_N$ ($\mu$m)& $\tilde{w}$ (nm) & ${\cal E}_{\rm cond}$ (J/m$^3$)& $\omega$ \\
	     \hline
	     \hline
	     UGe$_2$ & 14200 & 15 & 1  & 2 & 4 & 13.6 & 40  & 0.039 \\   
	     \hline 
	     UCoGe & 51 & 15 & 1.2 & 0.15 & 2  & 45 & 29  & 2.36 \\   
	     \hline
	     URhGe & 3070 & 18 & 0.9 & 0.2 & 20 & 3450  & 30 & 6.5 $\times 10^{-4}$ \\   
	     \hline
		\end{tabular}
	\caption{Experimental values the magnetization energy $2 \pi M_0^2$, the coherence length $\xi(0)$, the magnetic penetration depth $\lambda(0)$, the half-thickness $L_z$ and the normal-state domain width $l_N$ in the compounds UGe$_2$~\cite{saxena,ureview,uge2,sakarya}, UCoGe~\cite{huy,ureview,ucoge,hykel} and URhGe~\cite{aoki2001,ureview,urhge,dolocan}. The effective domain wall thickness $\tilde{w}$ is obtained from the experimental value of $l_N$ and $L_z$. For an estimate of the condensation energy, we used the BCS formula ${\cal E}_{\rm cond} \approx 0.166 T_c \Delta C$ where $\Delta C$ is the specific heat jump at $T_c$. The parameter $\omega$ is calculated with definition~(\ref{eq:omega}).}
\label{tab:exp}
\end{table}

\end{widetext}

\begin{figure}
\begin{center}
\scalebox{0.8}{
 \includegraphics*{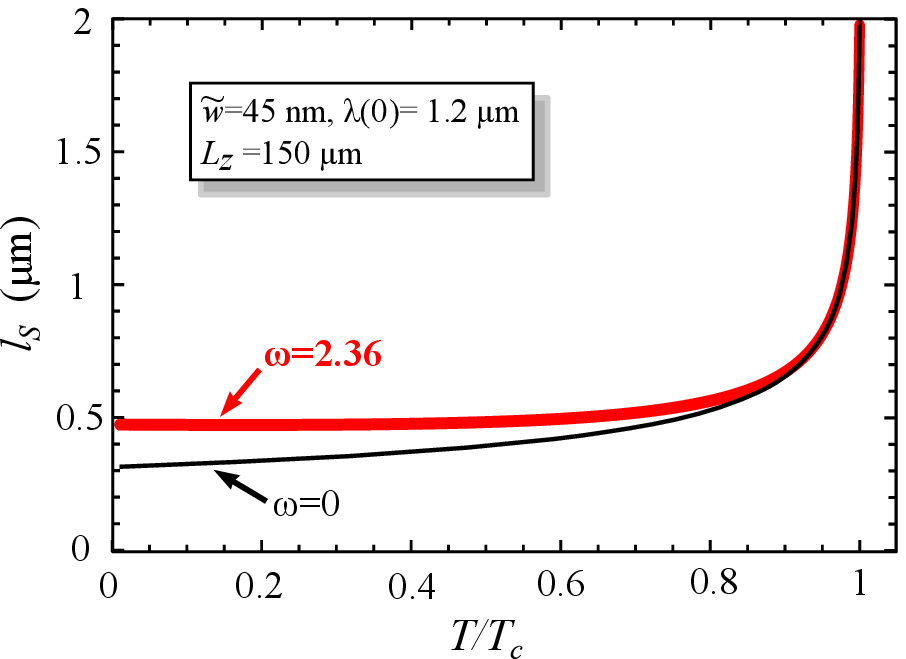}}
\end{center}
\caption{(color online). Temperature dependence of the domain width $l_S$ (thick line) calculated with experimental parameters for UCoGe in a film of half-thickness $L_z= 150$ $\mu$m. The thin line is $l_S(T)$ obtained with $\omega=0$, which corresponds to singlet superconductivity.}
\label{fig:ls_t-ucoge-1}
\end{figure}	

\begin{figure}
\begin{center}
\scalebox{.8}{
 \includegraphics*{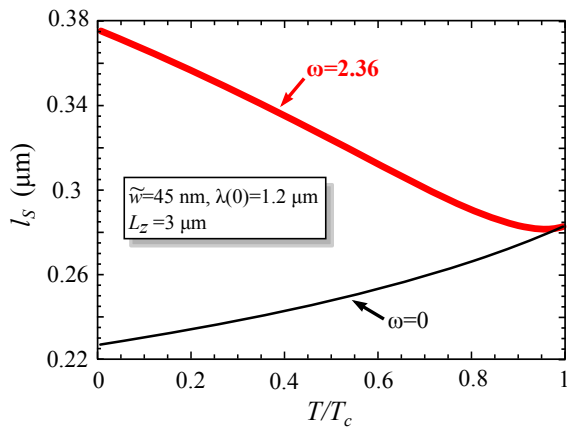}}
\end{center}
\caption{(color online). Temperature dependence of the domain width $l_S$ (thick line) calculated with experimental parameters for UCoGe in a film of half-thickness $L_z= 3$ $\mu$m. The thin line is $l_S(T)$ obtained with $\omega=0$, which corresponds to singlet superconductivity.}
\label{fig:ls_t-ucoge-2}
\end{figure}

According to experimental data (see Table~\ref{tab:exp}), the Uranium-based superconductors UGe$_2$, UCoGe and URhGe are in the limit $l_N \ll L_z$. The effective domain wall thickness $\tilde{w}$ can then be easily calculated in this limit from the experimental value of the domain width $l_N$ in the normal state. We roughly estimate the zero-temperature condensation energy with the BCS formula ${\cal E}_{\rm cond}(0) \approx 0.166 T_c \Delta C$ where $\Delta C$ is the volumic specific heat jump at $T_c$. Among the three compounds UCoGe has the smallest magnetization resulting into the largest $\omega \sim 1$ (see Table~\ref{tab:exp}). It is then the most promising candidate where to look for possible observation of the non-monotonic temperature dependence of the domain width in the superconducting phase. Using its estimated parameters we plotted the temperature dependence of the domain width for the singlet and the triplet scenario in a film of half-thickness $L_z= 0.15$ mm (Fig.~\ref{fig:ls_t-ucoge-1}) and $L_z=3 \mu$m (Fig.~\ref{fig:ls_t-ucoge-2}). For $L_z= 0.15$ mm (Fig.~\ref{fig:ls_t-ucoge-1}) the domain width decreases below $T_c$ in both the singlet and the triplet case but there is a significant difference of size at $T=0$ between the two. By reducing the film thickness the discrepancy is amplified. For instance with $L_z=3 \mu$m (Fig.~\ref{fig:ls_t-ucoge-2}) one obtains a qualitative difference: for the triplet case the width increases by 32\% from its value in the normal state while for the singlet case it shrinks by 21\%. 
	
\section{Conclusion}

We have done the complete analysis, within the London approximation, of the domain width $l_S $ at equilibrium in a SFM film of arbitrary thickness $2L_z$. We have shown that the ratio $l_S/L_z$ follows a universal dependence on the normalized penetration depth $\lambda/L_z$ and the normalized effective wall thickness $\tilde{w}/L_z$ (see Figs.~\ref{fig:ls01}-\ref{fig:ls03}). In addition to the exact numerical dependence, we have derived analytical expressions of this relation in limit cases. In particular, we have recovered the previously published results by Faur\'e and Buzdin~\cite{faure} which had been established for $L_z \gg \lambda$ and $l_S$, and we have complemented them with the analysis in the opposite limit $L_z \ll \lambda$ and $l_S$. 

We have found that the domain width always decreases with temperature when the DS enters the superconducting phase. The screening supercurrent induces a decrease of the domain width if the latter is larger than the penetration depth. However, the paramagnetic screening suppresses the DS below a critical value $\lambda_c$ of penetration depth, which means that the system may become mono-domain at a finite temperature if $\lambda(0) < \lambda_c$. The domain shrinkage is relatively small when the domain width $l_N$ in the normal state is much larger than the film thickness. Indeed, in this limit, the width decreases as $\lambda^2/L_z$ which is only a fraction smaller than $l_N$ since $\lambda_c \lesssim \sqrt{L_z l_N}$. With the other limit shape of domains, i.e. $l_N  \ll L_z$, the shrinkage can be important since then the domain width decreases as $\lambda$ while the critical penetration depth $\lambda_c  \sim l_N^2/L_z \ll l_N$.

Furthermore, we have investigated the effect of the triplet pairing within an effective temperature-dependent renormalization of the domain wall energy. Actually, the DS magnetization alternatively suppresses the order-parameter components with opposite spin-projections. This inhomogeneity results in an additional positive contribution to the domain wall energy, that is absent for the singlet pairing. The supplementary term vanishes at the transition temperature but can be significant at a lower temperature. The variation of the domain width is then no more described by the universal dependence of $l_S/L_z$ on $\lambda/L_z$ obtained for the singlet case. This gives a phenomenological way of distinguishing triplet from singlet pairing in experiments. And, as we have shown, the discrepancy can be amplified by reducing the film thickness (see Figs.~\ref{fig:ls_t-ucoge-1} and \ref{fig:ls_t-ucoge-2}).  

We have also established that vortices can appear when the magnetization exceeds a critical value $4\pi M_v$. In narrow domains (i.e. $l \ll L_z$), the latter is equal to the lower critical field $H_{c1}^*$ of a non-magnetic superconductor if $\lambda \ll l$, but in the opposite limit $\lambda \gg l$, it is much larger than $H_{c1}^*$. For wide domains (i.e. $l \gg L_z$), we have found that the critical magnetization behaves similarly, with the difference that $\lambda$ is then replaced by Pearl's effective penetration depth $\lambda_{\rm eff}$.

The available experimental data (Table~\ref{tab:exp} and Ref.~\cite{saxena,ureview,uge2,sakarya,huy,ureview,ucoge,hykel,aoki2001,ureview,urhge,dolocan}) for UGe$_2$, URhGe and UCoGe show that in these compounds $l \ll L_z$ and the conditions for the vortices appearance (section \ref{sec:vortex}) are fulfilled. This means that the effects of domain shrinkage (considered in section \ref{sec:width} for the case where the vortices are absent) should be weakened. Another difficulty to experimentally observe the evolution in the superconducting state may be related with the pinning of the domain wall and/or vortex pinning. At present there are no convincing experimental data on the change of the domain structure below $T_c$.

It has been recently observed in S/F bilayers~\cite{tamegai,vlasko-vlasov} that the coupling between vortices and magnetic domains leads to a strong shrinkage of domains in the presence of the oscillating field used for equilibration of the domain structure. It would be interesting to perform the similar experiments with superconducting ferromagnets to study these coupling effects. 

We thank K. Hasselbach, C. Paulsen, J.-P. Brison, and D. Aoki for helpful discussions. This work has been supported by the French ANR project SINUS.

\appendix

\section{The function $I(\Lambda)$}
We define the function $I(\Lambda)$ as the integral
\begin{equation}
 I(\Lambda) \equiv \int_{0}^{\infty} \!\!\!\!\! \frac{u \; du}{\sqrt{\frac{1}{\Lambda^{2}}+u^2}^3  \left( \sqrt{\frac{1}{\Lambda^{2}}+u^2} + u \coth \sqrt{\frac{1}{\Lambda^{2}}+u^2} \right)}.
\label{eq:I} 
\end{equation}

\subsection{Limit $\Lambda \gg 1$}
The integral can be rewritten as
\begin{eqnarray}
& I(\Lambda)  =  \int_{0}^{\infty} \!\!\!\!\! \frac{u \; du}{\left( \Lambda^{-2}+u^2 \right) \left(  \Lambda^{-2} + u^2 + u \right)} \nonumber \\
& + \int_{0}^{\infty}  \frac{u^2 \left( 1 - \sqrt{\Lambda^{-2} + u^2} \coth \sqrt{\Lambda^{-2} + u^2} \right)  du }{\left( \Lambda^{-2}+u^2 \right) \left( \Lambda^{-2} + u^2 + u \right) \left( \Lambda^{-2} + u^2 + u \sqrt{\Lambda^{-2} + u^2} \coth \sqrt{\Lambda^{-2} + u^2}  \right)},  \nonumber 
\end{eqnarray}
so by using the approximations $\sqrt{\Lambda^{-2} + u^2} \coth \sqrt{\Lambda^{-2} + u^2}  \approx  1 + (\Lambda^{-2} + u^2)/3$ in the numerator and $\approx 1$ in the denominator when $u<1$, and $\sqrt{\Lambda^{-2} + u^2} \coth \sqrt{\Lambda^{-2} + u^2} \approx u$ when $u>1$,
\begin{eqnarray}
 I(\Lambda) \approx & \int_{0}^{\infty} \!\!\!\!\! \frac{u \; du}{\left( \Lambda^{-2}+u^2 \right) \left( \Lambda^{-2} + u^2 + u \right)}  - \frac{1}{3} \int_0^1 \frac{u^2 \; du}{\left( \Lambda^{-2} + u^2 + u \right)^2} \nonumber \\
& + \int_1^{\infty} \frac{u^2 (1-u) du}{\left( \Lambda^{-2}+u^2 \right)  \left( \Lambda^{-2}+u^2 + u \right) \left( \Lambda^{-2}+ 2 u^2 \right)}. \nonumber
\end{eqnarray}

 Then for $\Lambda \gg 1$ 
\begin{equation}
I(\Lambda) \approx \frac{\pi \Lambda}{2} - 2 \ln \Lambda - \frac{11}{12} + \ln 2~.
\label{eq:F201}
\end{equation}
 
\subsection{Limit $\Lambda \ll 1$} 
In the limit $\Lambda \ll 1$, $\coth\sqrt{\Lambda^{-2} + u^2} \approx 1$ then
\begin{eqnarray}
I(\Lambda) & \approx & \int_0^{\infty} \frac{ u \; du}{\sqrt{\frac{1}{\Lambda^{2}}+u^2}^3  \left( \sqrt{\frac{1}{\Lambda^{2}}+u^2} + u  \right)} \nonumber \\ 
& = & \Lambda^2 \int_0^{\infty} \frac{ v \; dv}{\sqrt{1+v^2}^3  \left( \sqrt{1+v^2} + v \right)} \nonumber \\
& \approx & (1-\ln 2) \Lambda^2~.
\label{eq:F202}
\end{eqnarray}

\end{document}